\documentclass[aps,twocolumn]{revtex4-1}
\usepackage{amsmath,amssymb,graphicx,enumerate}

\newcommand{\amp}{&}
\newcommand{\mpl}{M_{Pl}}
\newcommand{\beq}{\begin{equation}}
\newcommand{\eeq}{\end{equation}}
\newcommand{\bea}{\begin{eqnarray}}
\newcommand{\eea}{\end{eqnarray}}
\newcommand{\bp}{\lambda}
\newcommand{\dr}{\Gamma_\phi}
\newcommand{\drreq}{\Gamma_{\phi,\mbox{\tiny{req}}}}
\newcommand{\as}{A}
\newcommand{\Nphi}{N_\phi}
\newcommand{\Nphib}{N_{\bar{\phi}}}
\newcommand{\nphi}{n_\phi}
\newcommand{\nphib}{n_{\bar{\phi}}}
\newcommand{\Nb}{N_b}
\newcommand{\Nbb}{N_{\bar{b}}}

\newcommand{\bphi}{b_\phi}
\newcommand{\gamf}{\gamma}
\newcommand{\fud}{\beta}

\begin{document}

\title{Baryogenesis from the Inflaton Field}

\begin{abstract}
In this letter we show that the inflaton can generate the cosmological baryon asymmetry. We take the inflaton to be a complex scalar field with a weakly broken global symmetry and develop a new variant on the Affleck-Dine mechanism. The inflationary phase is driven by a quadratic potential whose amplitude of B-modes is in agreement with BICEP2 data.
We show that a conserved particle number is produced in the latter stage of inflation, which can later decay to baryons. We present promising embeddings in particle physics, including the use of high dimension operators for decay or using a colored inflaton. We also point out  observational consequences, including a prediction of isocurvature fluctuations, whose amplitude is just below current limits, and a possible large scale dipole.
\end{abstract}

\author{Mark P.~Hertzberg$^*$ and Johanna Karouby$^\dagger$}
\affiliation{Center for Theoretical Physics and Dept.~of Physics\\
Massachusetts Institute of Technology, Cambridge, MA 02139, USA}

\date{\today}

\vspace*{-14.2cm} {\hfill MIT-CTP 4485\,\,\,\,} 

\maketitle

{\em Introduction}.---One of the outstanding challenges of modern particle physics and cosmology is to explain the asymmetry between matter and anti-matter throughout the universe. This asymmetry is quantified by the baryon-to-photon ratio $\eta$, which shows an over-abundance of matter at the level of $\eta_{obs}\approx6\times10^{-10}$, as measured in \cite{Ade:2013uln}.
One might try to dismiss this problem by assuming the universe simply began with the asymmetry. However, such a proposal appears both unsatisfying and unlikely due to cosmological inflation; a phase of exponential expansion in the early universe that helps to explain the large scale homogeneity, isotropy, and flatness, as well as the density fluctuations  \cite{Guth}. Such a phase would wipe out any initial baryon number. It is usually thought that this requires new fields to enter after inflation in the radiation (or matter) eras to generate the asymmetry (for reviews see  \cite{Riotto:1998bt}), such as at the electroweak phase transition (e.g., see \cite{Trodden:1998ym}). Since we have yet to see new physics beyond the Standard Model at the electroweak scale \cite{ATLAS}, it is entirely possible that baryogenesis is associated with much higher energies, and inflation is a probe into these high scales.

In this letter, and accompanying paper \cite{BaryogenesisPaper}, we show that although inflation wipes out any initial matter/anti-matter asymmetry, the asymmetry can still be generated by the inflaton itself. The key reason this is possible is that the inflaton acquires a type of vev during inflation and this information is not wiped out by the inflationary phase. In order to connect this to baryogenesis, we will put forward a new variation on the classic Affleck-Dine \cite{AffleckDine} mechanism for baryogenesis, which uses scalar field dynamics to obtain a net baryon number. In the original proposal, Affleck-Dine used a complex scalar field, usually thought to be unrelated to the inflaton but possibly a spectator field during inflation, to generate baryons in the radiation or matter eras. Various versions, often including connections to supersymmetry, have been found for these Affleck-Dine models, e.g.,  see \cite{Allahverdi:2012ju}.

In this letter we propose a new model where the aforementioned complex scalar field is the inflaton itself. 
In the accompanying paper \cite{BaryogenesisPaper}, we develop and provide details of this proposal, including both particle physics and cosmological aspects, and discuss current observational constraints. 
Our key ideas and findings are summarized as follows:
We propose that the inflaton is a complex scalar field with a weakly broken global $U(1)$ symmetry. 
For simplicity, we consider inflation driven by a symmetric quadratic potential, plus a sub-dominant symmetry breaking term.
The quadratic potential establishes tensor modes in agreement with recent BICEP2 results \cite{Ade:2014xna}. 
Given these recent cosmological observations, it is very important to establish a concise, predictive model as we do here.
We show that a non-zero particle number is generated in the latter stage of inflation. After inflation this can decay into baryons and eventually produce a thermal universe.
We propose two promising particle physics models for both the symmetry breaking and the decay into baryons:
(i) Utilizing high dimension operators for decay, which is preferable if the inflaton is a gauge singlet.
(ii) Utilizing low dimension operators for decay, which is natural if the inflaton carries color. 
We find that model (i) predicts the observed baryon asymmetry if the decay occurs through operators controlled by $\sim$\,GUT scale and this is precisely the regime where the EFT applies,
while model (ii) requires small couplings to obtain the observed baryon asymmetry. 
We find a prediction of baryon isocurvature fluctuation at a level consistent with the latest CMB bounds, which is potentially detectable.

In summary, our new results beyond the existing literature includes: 
(a) the direct comparison to the latest data; this includes the latest bounds on tensor modes, scalar modes, and baryon asymmetry, (b) the development of a broad framework to identify inflation with the origin of baryon asymmetry, without the detailed restrictions of supersymmetry, (c) specific model building examples including the cases of a singlet inflaton and a colored inflaton,
(d) predictions for isocurvature modes and compatibility with existing bounds, while standard Affleck-Dine models are ruled out if high-scale inflation occurred, (e) predictions of a large scale dipole.

{\em Complex Scalar Model}.---Consider a complex scalar field $\phi$, with a canonical kinetic energy $|\partial\phi|^2$, minimally coupled to gravity, with dynamics governed by the standard two-derivative Einstein-Hilbert action.
Our freedom comes from the choice of potential function $V(\phi,\phi^*)$. It is useful to decompose the potential into a ``symmetric" piece $V_s$ and a ``breaking" piece $V_b$ piece, with respect to a global $U(1)$ symmetry $\phi\to e^{-i\alpha}\phi$, i.e.,
$
V(\phi,\phi^*)=V_s(|\phi|)+V_b(\phi,\phi^*).
$
In order to describe inflation we assume that the symmetric piece $V_s$ dominates, even at rather large field values where inflation occurs.
For simplicity, we take the symmetric piece to be quadratic
$V_s(|\phi|)=m^2|\phi|^2$.
It is well known that a purely quadratic potential will establish large field, or ``chaotic" inflation \cite{Linde83}.
This is a simple model of inflation that will provide a useful pedagogical tool to describe our  mechanism for baryogenesis. Such a model is in good agreement with the spectrum of density fluctuations in the universe \cite{Ade:2013uln}, it is in agreement with the measured tensor modes from BICEP2 data \cite{Ade:2014xna}, and is motivated by simple symmetry arguments \cite{Hertzberg:2014aha}. Generalizing to other symmetric potentials is also possible.

The global symmetry is associated with a conserved particle number. So to generate a non-zero particle number (that will decay into baryons) we add a higher dimension operator that explicitly breaks the global $U(1)$ symmetry
$V_b(\phi,\phi^*)=\bp(\phi^n+\phi^{*n})$,
with $n\ge3$. 
We assume that the breaking parameter $\bp$ is very small so that the global symmetry is only weakly broken.
This assumption of very small $\bp$ is motivated by two reasons:
Firstly, since $\bp$ is responsible for the breaking of a symmetry, it is technically natural for it to be small according to the principles of effective field theory.
Secondly, the smallness of $\bp$ is an essential requirement on any inflationary model so that such higher order corrections do not spoil the flatness of the potential $V_s$. We also note that our model carries a discrete $\mathbb{Z}_n$ symmetry that makes it radiatively stable.

\begin{figure}
\center{\includegraphics[width=4.7cm]{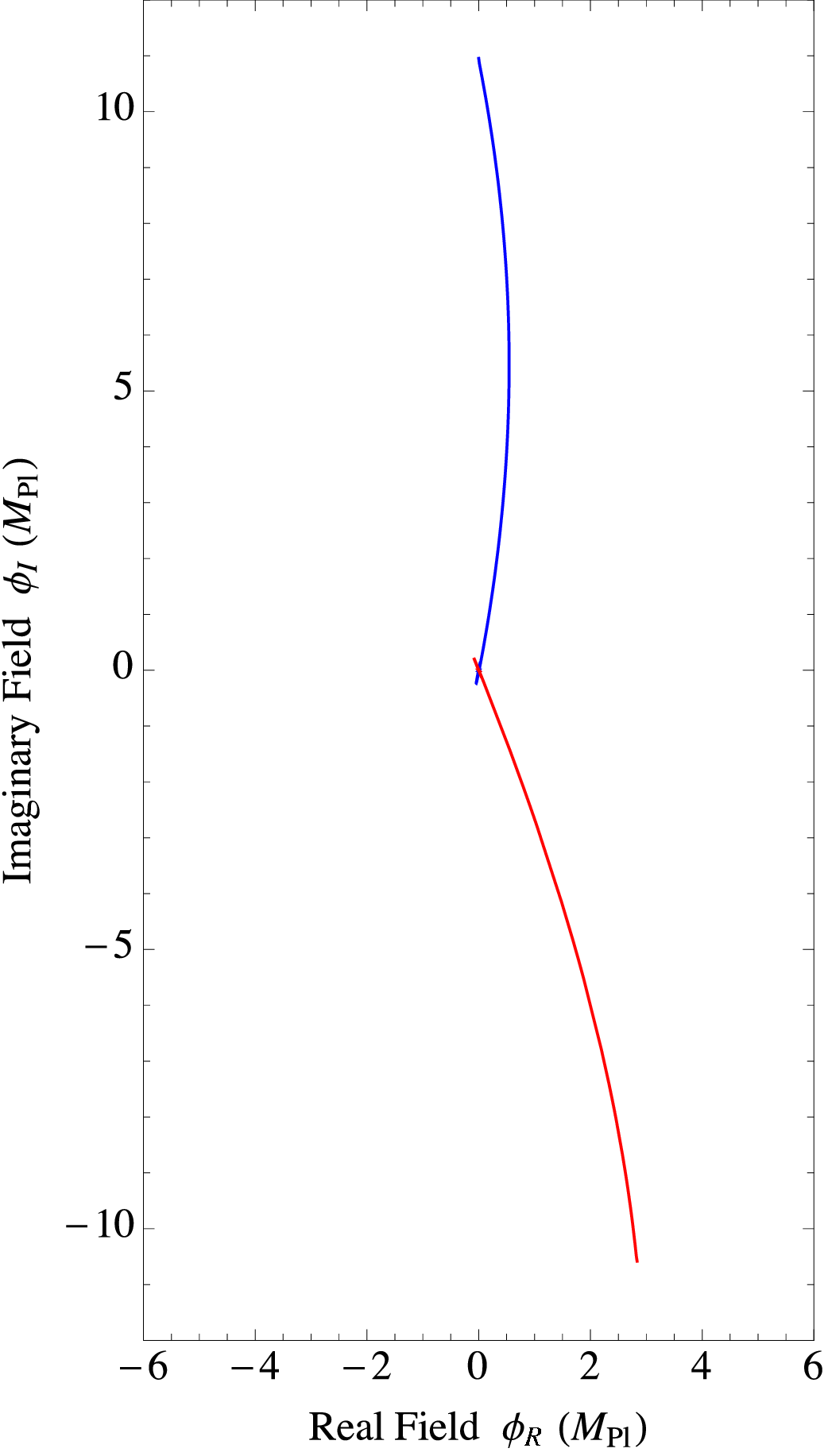}
\includegraphics[width=3.65cm]{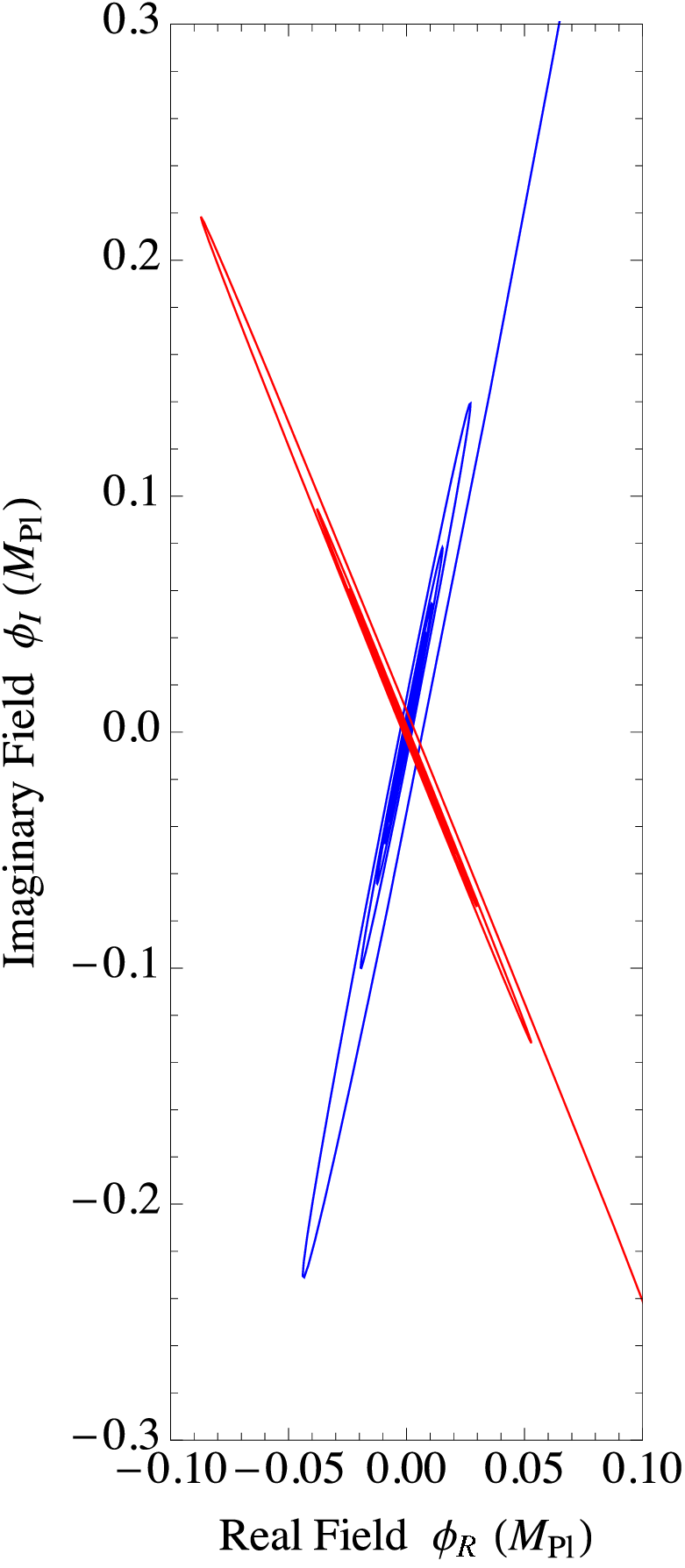}}
\caption{Field evolution in the complex $\phi$-plane for $n=3$ and $\bp\mpl/m^2=0.006$, with initial condition $\rho_i=2\sqrt{60}\,\mpl$. Left is zoomed out and shows early time behavior during slow-roll inflation. Right is zoomed in to $\phi=0$ and shows late time elliptic motion. Blue (upper) curve is for initial angle $\theta_i=\pi/2$ and red (lower) curve is for initial angle $\theta_i=-5\pi/12$.}
\label{FieldEvolution}\end{figure}

{\em Particle/Anti-Particle Asymmetry}.---We assume the field begins at large field values ($|\phi|\gg\mpl$) and drives inflation.
The field exhibits usual slow-roll and then redshifts to small values at late times, where it exhibits elliptic motion. This evolution is seen in Fig.~\ref{FieldEvolution} for two different initial conditions. Since $n\geq3$, then at late times the inflaton $\phi$ becomes small, the $\phi\to e^{-i\alpha}\phi$ symmetry violating term becomes negligible, and the symmetry becomes respected. By Noether's theorem this is associated with a conserved particle number. In an FRW universe with scale factor $a(t)$ and comoving volume $V_{com}$, this is
\beq
\Delta \Nphi=\Nphi-\Nphib=i\, V_{com}\,a^3(\phi^*\dot\phi-\dot\phi^*\phi).
\eeq
To be self consistent we ignore spatial gradients, and the equation of motion for $\phi$ is:
$
\ddot\phi+3 H\dot\phi+m^2\phi+\bp\,n\,\phi^{* n-1}=0,
$
where $H=\dot{a}/a$ is the Hubble parameter.

For small $\bp$ we can reduce the complexity of the problem significantly. By using the equation of motion, we can obtain an integral expression for $\Delta\Nphi$ which is proportional to $\bp$. This allows us to compute the evolution of the field to zeroth order in $\lambda$, which implies radial motion in the complex plane. We rewrite the zeroth order motion of the field in polar co-ordinates as $\phi_0(t)=e^{i\theta_i}\rho(t)/\sqrt{2}$, where $\theta_i$ is the initial angle of the field at the beginning of inflation.
The problem then reduces to solving only a single ordinary differential equation. At first order in $\bp$, $\Delta \Nphi$ is simply
\beq
\Delta \Nphi(t_f) = - \bp{V_{com}\,n\over 2^{{n\over2}-1}}\sin(n\,\theta_i)\int _{t_i}^{t_f} dt\,a(t)^3\rho_0(t)^n,
\label{deltaNsimp}\eeq
Here $\rho_0$ is a real valued function satisfying the quadratic potential version of the equation of motion
$
\ddot\rho_0+3 H_0\dot\rho_0+m^2\rho_0=0,
$
with corresponding Friedmann equation (we assume flat FRW)
$
H_0^2=\varepsilon_0/3\mpl^2$ and energy density $\varepsilon_0=\dot\rho_0^2/2+m^2\rho_0^2/2,
$
where $\mpl\equiv1/\sqrt{8\pi G}$ is the reduced Planck mass. So by solving for a single degree of freedom in a quadratic potential, we have an expression for the particle number in the small $\bp$ regime. 
We note that for particular values of the initial angle $\theta_i$, such that $\theta_i={p\pi\over n}\,|\,p\in\mathbb{Z}$, no asymmetry is generated due to the $\sim\sin(n\,\theta_i)$ factor. Since we are interested in baryogenesis, we consider $\theta_i$ to be a typical generic value rather than these special ones.

\begin{figure}
\center{\includegraphics[width=\columnwidth]{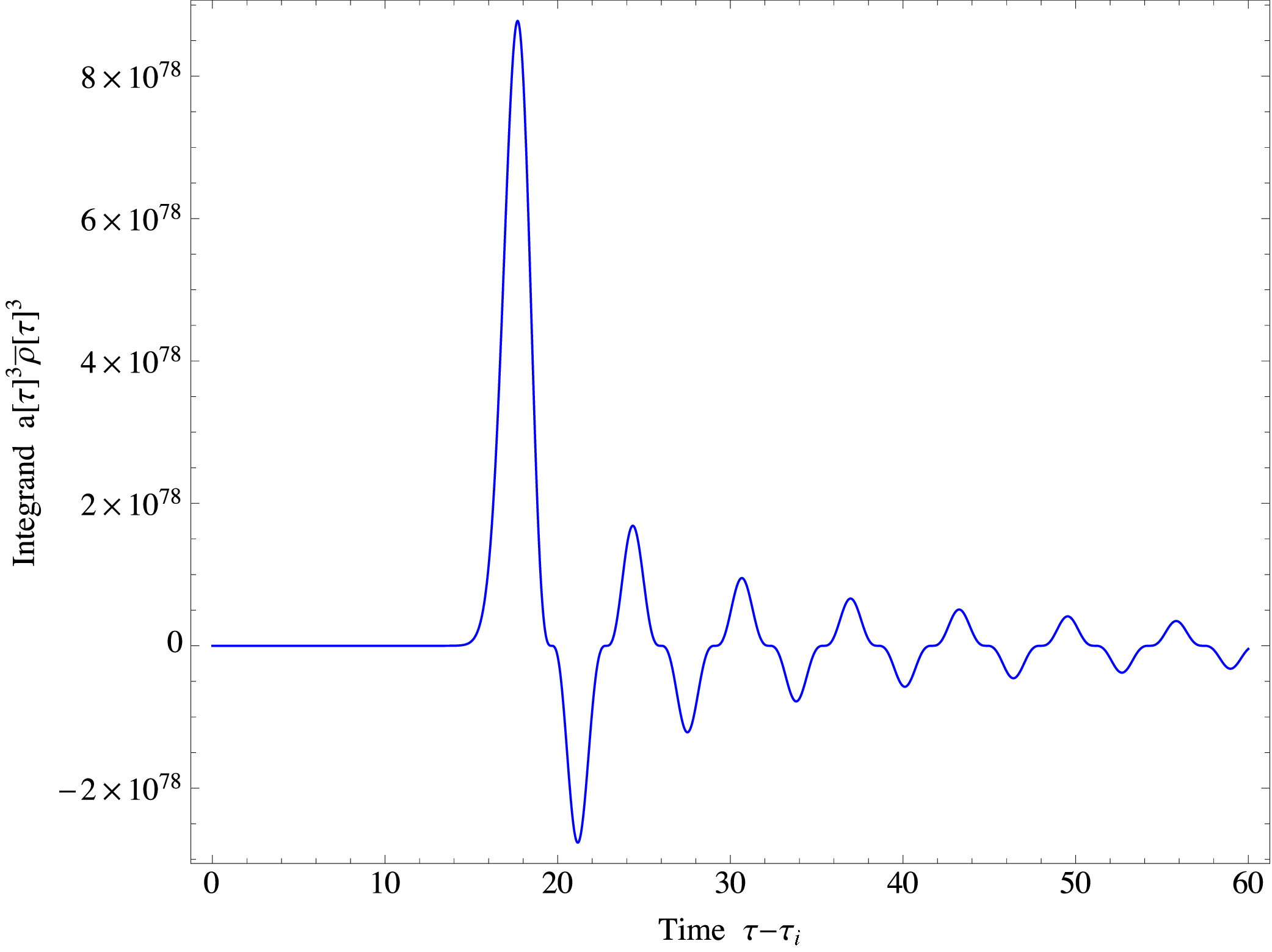}}
\caption{The integrand giving $\Delta \Nphi$ (eq.~(\ref{deltaNsimp})), with respect to dimensionless variables $\tau=mt$, $\bar{\rho}=\rho_0/\mpl$. In this plot we have taken $n=3$ and initial conditions $\bar{\rho}_i=2\sqrt{60}$, $a_i=1$. The large peak is in the latter phase of inflation; so this is where most of the $\phi$ (or anti-$\phi$) particles are produced.}
\label{Integrand}\end{figure}

The integrand in eq.~(\ref{deltaNsimp}) is plotted in Fig.~\ref{Integrand} using dimensionless variables $\tau\equiv mt$ and $\bar{\rho}\equiv\rho_0/\mpl$.
In the limit in which we take $\tau_i$ very early during slow-roll inflation and we take $\tau_f$ very late after inflation, then the integral in eq.~(\ref{deltaNsimp}) becomes independent of both $\tau_i$ and $\tau_f$. 
The dominant contribution to the integral, and in turn the dominant production of $\phi$ particles (or anti-particles) occurs in the latter stage of inflation.
This is nicely seen in Fig.~\ref{Integrand}. It can be shown that for the parameters of the figure, the end of inflation is $\tau\approx18$, which is precisely at the end of the sharp rise and fall of the integrand. This is shifted to slightly earlier times for higher $n$. 

{\em Dimensionless Asymmetry}.---Although $\Delta \Nphi$ is dimensionless, it is extrinsic, depending on the size of the universe. It is useful to define a related intrinsic quantity, which provides a measure of the asymmetry
$
\as\equiv\Delta \Nphi/(\Nphi+\Nphib).
$
The denominator of $\as$ can be related to the energy density stored in the field, because after inflation $\phi$ is effectively a gas of non-relativistic $\phi$ and anti-$\phi$ particles with energy density
$\varepsilon_0 = m (\nphi+\nphib)$. We find that this asymmetry parameter takes on the simple form
\beq
\as = -c_n{\bp\mpl^{n-2}\over m^2}\sin(n\,\theta_i).
\label{Rapprox}\eeq
Numerically solving the dimensionless ordinary differential equation for $\rho_0$ and then integrating, leads to the following results for the coefficient $c_n$ for the first few $n$
\bea
&&c_3\approx 7.0,\,\,\,c_4\approx 11.5,\,\,\,c_5\approx 14.4,\,\,\,c_6\approx 21.8,\nonumber\\
&&c_7\approx 34.8,\,\,\,c_8\approx59.3,\,\,\,c_9\approx107,\,\,\,c_{10}\approx 201.\,\,\,\,
\label{cnvalues}\eea
In our companion paper \cite{BaryogenesisPaper} we prove that for high $n$, the coefficients are given by
\beq
c_n\approx \tilde{c}\,2^{n/2}3^{-n/2}\,n \,\Gamma_{1\over2}(n/2),
\label{cappx2}\eeq
where $\tilde{c}$ is a coefficient given by $\tilde{c}\approx 6.64$ and $\Gamma_a$ is the incomplete gamma function. We find this result to be surprisingly accurate even for small $n$.

{\em Baryon Asymmetry}.---Recall that the baryon asymmetry is defined as the ratio of baryon difference to photon number at late times
$
\eta\equiv(\Nb-\Nbb)_f/(N_\gamma)_f,
$
where $f$ indicates the late time, or ``final" value, after decay and thermalization.
We associate with each $\phi$ particle a baryon number $\bphi$; for instance $\bphi=1$ or $\bphi=1/3$ in simple models. We assume that the decay of $\phi$ and all subsequent interactions is baryon number conserving, so we can relate the final number to the initial number as follows:
$(\Nb-\Nbb)_f=\bphi(\Nphi-\Nphib)_i$,
where $i$ indicates the early time, or ``initial" value, before decay and thermalization (but well after the baryon violating processes have stopped).     

At early times we can relate the number of $\phi$ particles to the Hubble parameter as
$(\Nphi+\Nphib)_i=3\mpl^2V_{com}(a^3 H^2)_i/m$.
At late times we can relate the number of photons to the temperature as
$(N_\gamma)_f = 2\zeta(3)V_{com}(a^3 T^3)_f/\pi^2$.
By assuming that the thermalization is rapid, we can simply evaluate both the ``initial" and ``final" quantities around the time of decay.
Denoting the decay rate of $\phi$ as $\dr$, then thermalization occurs around $H\approx \dr$ \cite{Kofman:1997yn}, which allows us to solve for the reheat temperature in terms of the number of relativistic degrees of freedom $g_*$. Putting all this together gives the following expression for the baryon-to-photon ratio
\beq
\eta\approx {\fud\,\pi^{7/2}g_*^{3/4}\bphi\over 2^{7/4}3^{1/2}5^{3/4}}{\as\,\dr^{1/2}\mpl^{1/2}\over m},
\eeq
where $\fud$ is an $\mathcal{O}(1)$ fudge factor from the details of the transition from the $\phi$ era to the thermal era.

{\em Constraints from Inflation}.---An important constraint is that the symmetry breaking term in the potential $\bp(\phi^n+\phi^{*n})$ be subdominant during inflation. Since this contribution to the potential goes negative at large field values, we obviously need it to be small during inflation. 
For quadratic inflation, it can be shown that the field value is $\rho_i\approx2\sqrt{N_e}\,\mpl$ during inflation; this leads to the constraint
\beq
\bp\ll\bp_0\equiv{m^2\over 2^{n/2}N_e^{n/2-1}\mpl^{n-2}}.
\label{gammacons}\eeq
We now use the threshold value $\bp_0$ and the above set of equations to impose a condition on the decay rate in order to obtain $\eta_{obs}$
\bea
\drreq\amp\approx\amp 10^{-7}\,\mbox{eV}\times 2^{n+1}N_e^{n-2}c_n^{-2}   \left(\bp_0\over\bp\right) \nonumber\\
\amp\times\amp\left(m\over 10^{13}\,\mbox{GeV}\right)^{\!2} \Bigg{(} {g_*^{3/4}\over 30}\fud\, \bphi|\sin(n\,\theta_i)|\Bigg{)}^{\!-2}\!.\,\,\,\,
\label{drcond}\eea
To provide concrete quantitative results for the required decay rate, we assume that the coupling $\bp$ is a factor of 10 smaller than its inflationary upper bound $\bp_0$, $\fud \bphi|\sin(n\,\theta_i)|\approx 1$, $m\approx 1.5\times 10^{13}$\,GeV (required for the correct amplitude of fluctuations in quadratic inflation), $N_e\approx 55$, $g_*\approx 10^2$, and we insert the $c_n$ from eq.~(\ref{cnvalues}). We find that $\drreq$ increases with $n$; a few examples are
\bea
&&n=3\Rightarrow\drreq\approx 4\times 10^{-5}\,\mbox{eV},\nonumber\\
&&n=4\Rightarrow\drreq\approx 2\times 10^{-3}\,\mbox{eV},\nonumber\\
&&n=8\Rightarrow\drreq\approx 9\times 10^3\,\mbox{eV}\nonumber\\
&&n=10\Rightarrow\drreq\approx 10^7\,\mbox{eV}.
\label{drvalues}\eea
In all these cases, the corresponding reheat temperature is much bigger than $\sim$\,MeV, the characteristic temperature of big bang nucleosynthesis (BBN).  We now examine to what extent these decay rates can be realized in two particle physics models.

{\em High Dimension Operators}.---In the simplest case one can take $\phi$ to be a gauge singlet. 
Without further refinement, this would allow $\phi$ to decay into non-baryonic matter, such as Higgs particles, through operators such as as $\sim\phi\, H^\dagger H$.
A natural way around this problem is to suppose that the global $U(1)$ symmetry is almost an exact symmetry of nature (or at least in the $\phi$ sector). 
Of course global symmetries cannot be exact. If nothing else, they must be broken by quantum gravity effects. Usually this implies the breaking of the symmetry by some high dimension operator. For high $n$, the breaking parameter will need to satisfy $\bp\lesssim(\mbox{few}/\sqrt{G})^{4-n}$ to be be consistent with inflation. This is compatible with quantum gravity expectations.
Another way to argue this is simply to impose a discrete $\mathbb{Z}_n$ symmetry.

So we can either imagine that the $U(1)$ symmetry breaking occurs at dimension $n\ge8$ operators or only operators that respect the $\mathbb{Z}_n$ symmetry. Then all low dimension operators that break the symmetry, such as $\sim\phi\, H^\dagger H$, would be forbidden. 
Since $\phi$ carries baryon number, then up to dimension 7 it could only decay into quarks through operators of the form
\beq
\Delta\mathcal{L}\sim{c\over \Lambda^3}\phi^*q\,q\,q\,l+\mbox{h.c},
\eeq
where we are suppressing indices for brevity.
Here we have introduced an energy scale $\Lambda$ that sets the scale of new physics (and the cutoff on the field theory) and $c$ is some dimensionless coupling.

At large amplitude, it is possible for parametric resonance to occur \cite{Kofman:1997yn}. However, one can find a sensible parameter regime where standard perturbative decay rates apply. We shall assume this here, and leave the other regime for future work \cite{HertzbergSelfResonance}.
The perturbative decay rate associated with this operator is roughly
\beq
\dr(\phi\to q+q+q+l)\sim {c^2\over 8\pi} {m^7\over \Lambda^6}.
\eeq
We now compare this to the required decay rates from eq.~(\ref{drvalues}).
For $m\approx 1.5\times 10^{13}$\,GeV and $c=\mathcal{O}(1)$, we find that the model has the required decay rate for
$\Lambda$ in the range $10^{15}-10^{16}\,\mbox{GeV}$,
for $n=8,\,10,\,12$, which is intriguingly around the GUT scale.
Also, since this scale satisfies: $H_i\ll\Lambda\ll\mpl$, then this is precisely within the regime of validity of the EFT.

On the other hand, lower values of $n$ do have their own advantage: They tend to lead lower values of the reheat temperature, which may be relevant to avoid potential problems with sphaleron washout.

{\em Colored Inflaton}.---Another possibility is to allow the inflaton to carry color. 
So lets give $\phi$ a color index, $i=r,w,b$, and allow for ``up" $\phi_u$ and ``down" $\phi_d$ versions and different generations labelled by $g$.
We can construct $U(1)$ violating terms in the potential that respect the $SU(3)_c$ symmetry. For instance, at dimension $n=3$, we can introduce the breaking term
\beq
V_b(\phi,\phi^*) = \bp^{gg'\!g''}  \varepsilon_{ii'\!i''} \, \phi_{ug}^{i}\,\phi_{dg'}^{i'} \, \phi_{dg''}^{i''}+\mbox{h.c},
\eeq
where $\varepsilon_{ijk}$ is the totally anti-symmetric tensor, and we have summed over color indices and different generations. 
This is the leading $U(1)$ violating operator, but this can be generalized to higher operators. We note that we are not especially sensitive to corrections from gluons due to asymptotic freedom \cite{GrossWilczek}.

Since $\phi$ carries color, we can readily build operators that mediate $\phi$ decay into quarks, while respecting the global symmetry, such as the following dimension 4 operator
$
\Delta\mathcal{L}\sim y\,\phi^{i*} q^i \bar{f}+\mbox{h.c},
$
where $f$ is some color neutral fermion and $y$ is a type of Yukawa coupling. 
This decay rate is roughly
$
\dr(\phi\to q+\bar{f})\sim y^2m/8\pi.
$
For high scale inflation, such as quadratic inflation that we discussed earlier, the inflaton mass is large $m\sim 10^{13}$\,GeV, so one would require an extremely small value of $y$ to obtain decay rates comparable to the required values we computed earlier in eq.~(\ref{drvalues}). 
In certain settings, such as supersymmetry (which would provide extra motivation for the existence of such colored scalars, or ``squarks"), one could examine if some non-renormalization theorem may help to stabilize $y$ at such small values. 
Another possibility would be to take $\bp$ much smaller than $\bp_0$, which would allow for higher values of $y$.

In the case of a colored inflaton, one would like an explanation as to why the inflationary potential is sufficiently flat for inflation to occur. A charged inflaton will tend to lead to loop corrections that steepen the potential. Though this is potentially avoidable.

{\em Isocurvature Fluctuations}.---Quantum fluctuations from inflation provide an excellent candidate for the origin of density fluctuations in the universe.
In simple single field models, only a curvature (``adiabatic") fluctuation is generated, due to fluctuations in the inflaton. 
For multi-field inflationary models, an isocurvature (``entropic") fluctuation is also generated \cite{Bartolo:2001rt}. This is due to quantum fluctuations in the field orthogonal to the classical field trajectory, which leaves the total density unchanged. 
Here the complex (two field) model will generate (baryonic) isocurvature fluctuations in $\phi$.

In our companion paper \cite{BaryogenesisPaper} we derive the following ratio of isocurvature fluctuations to total (primarily adiabatic) fluctuations in the CMB
\beq
\alpha_{II}\approx {32\gamf^2\over5}{\Omega_b^2\over\Omega_m^2}{n^2\mpl^2\epsilon_{sr}\over\rho_i^2}\cot^2(n\,\theta_i).
\label{alpha1}\eeq
where $\gamma=\mathcal{O}(1)$ from de Sitter random walk and $\epsilon_{sr}$ is the first slow-roll parameter.
Planck data reveals that the baryon-to-matter ratio is $\Omega_b/\Omega_m\approx 0.16$. Lets take $\gamf\sim2$, $\cot(n\,\theta_i)\sim1$, and specialize to the case of quadratic inflation with $\epsilon_{sr}\approx1/(2 N_e)$ and $\rho_i\approx2\sqrt{N_e}\,\mpl$, in agreement with BICEP2 data \cite{Ade:2014xna}. Then setting $N_e\approx 55$, we have our prediction for the isocurvature fraction
\beq
\alpha_{II}\sim 3\times 10^{-5}\,n^2.
\eeq
Recent Planck results have provided an upper bound on cold dark matter isocurvature fluctuations of \cite{Ade:2013uln}
$\alpha_{II}<3.9\times 10^{-2}$ at 95\% confidence,
and we shall use this as a rough bound on baryon isocurvature fluctuations.
For the lowest value of $n$, namely $n=3$, we predict $\alpha_{II}\sim 3\times 10^{-4}$, i.e., two orders of magnitude below the current bound. 
For moderately high value of $n$, such as $n= 8,\,10,\,12$ (as motivated earlier), then our prediction is $\alpha_{II}\sim 3\times 10^{-3}$, i.e., only one order of magnitude below the current bound. This is quite exciting as it is potentially detectable in the next generation of data.

{\em Discussion}.---1. In this letter we have proposed a way to directly unify early universe inflation and baryogenesis, with motivation from the Affleck-Dine mechanism.
These models intertwine parameters of high energy particle physics and inflation in an interesting way 
(for related discussions, see \cite{Hertzberg:2011rc}).
For instance, for the colored inflaton model, higher values of $m$ are preferred in order to obtain $\eta_{obs}$, so this favors high $H_i$ and appreciable tensor modes as seen by BICEP2 \cite{Ade:2014xna}.
While for the gauge singlet model, we found a connection to the $\sim$\,GUT scale.

2. Since $\eta\propto-\sin(n\,\theta_i)$, then for inhomogenous $\theta_i$, this leads to a large scale baryon dipole in the universe, which could be relevant to CMB anomalies \cite{Ade:2013nlj}. Also, it could lead to a multiverse of different baryon number.
Furthermore, the quantum isocurvature fluctuations are potentially detectable, but consistent with constraints. In more standard Affleck-Dine models, where $\phi$ is not the inflaton, the isocurvature fluctuation can be so large that many models are already ruled out. This can be seen from eq.~(\ref{alpha1}) due to the factor of $1/\rho_i^2$. For fields other than the inflaton, their vev's are typically sub-Planckian, thus leading to a huge isocurvature fluctuation. In contrast, for our inflaton models, a small, but potentially detectable, isocurvature fluctuation is natural.

3. Further work includes extension to other inflation models, further detailed embedding in particle physics, and to examine parametric resonance after inflation \cite{HertzbergSelfResonance}.

{\em Acknowledgments}.---We would like to thank Mustafa Amin, Michael Dine, Alan Guth, Jesse Thaler, Tanmay Vachaspati, and Frank Wilczek for discussion, and we would like to acknowledge support by the Center for Theoretical Physics at MIT. This work is supported by the U.S. Department of Energy under cooperative research agreement Contract Number DE-FG02-05ER41360. JK is supported by an NSERC PDF fellowship.

{\em Email}.---$^*$mphertz@mit.edu, $^\dagger$karoubyj@mit.edu

\end{document}